\documentclass[]{spie}  

\usepackage{amsmath,amsfonts,amssymb}
\usepackage{graphicx}
\usepackage[colorlinks=true, allcolors=blue]{hyperref}

\newcommand\dslm{Dragonfly Spectral Line Mapper}
\newcommand\dslmp{pathfinder Dragonfly Spectral Line Mapper}

\title{The pathfinder Dragonfly Spectral Line Mapper: Pushing the limits for ultra-low surface brightness spectroscopy}

\author[a]{Deborah M. Lokhorst}
\author[b,c]{Seery Chen}
\author[d]{Imad Pasha}
\author[b,c]{Jeff Shen}
\author[e]{Evgeni I. Malakhov}
\author[b,c]{Roberto G. Abraham}
\author[d]{Pieter van Dokkum}

\affil[a]{NRC Herzberg Astronomy \& Astrophysics Research Centre,
5071 West Saanich Road, 
Victoria, BC V9E2E7, Canada}
\affil[b]{David A. Dunlap Department of Astronomy \& Astrophysics,
University of Toronto,
50 St. George Street, 
Toronto, ON M5S3H4, Canada}
\affil[c]{Dunlap Institute,
University of Toronto,
50 St. George Street, 
Toronto, ON M5S3H4, Canada}
\affil[d]{Department of Astronomy,
Yale University,
52 Hillhouse Ave., New Haven, CT 06511, USA}
\affil[e]{New Mexico Skies, Inc., 
9 Contentment Crest, Mayhill, NM 88339, USA}

\authorinfo{Send correspondence to D.M.L.: deborah.lokhorst@nrc-cnrc.gc.ca}

\begin{document} 
\maketitle

\begin{abstract} 
The pathfinder Dragonfly Spectral Line Mapper is a distributed aperture telescope based off of the Dragonfly Telephoto Array with additional instrumentation (the Dragonfly ``Filter-Tilter'') to enable ultranarrow bandpass imaging. The pathfinder is composed of three redundant optical tube assemblies (OTAs) which are mounted together to form a single field of view imaging telescope (where the effective aperture diameter increases as the square-root of the number of OTAs). The pathfinder has been on sky from March 2020 to October 2021 equipped with narrowband filters to provide proof-of-concept imaging, surface brightness limit measurements, on sky testing, and observing software development. Here we describe the pathfinder telescope and the sensitivity limits reached along with observing methods. We outline the current limiting factors for reaching ultra-low surface brightnesses and present a comprehensive comparison of instrument sensitivities to low surface brightness line emission and other methods of observing the ultra-faint line emission from diffuse gas. Finally, we touch on plans for the upcoming 120-OTA Dragonfly Spectral Line Mapper, which is currently under construction.

\end{abstract}


\section{INTRODUCTION}
\label{sec:intro}  

Directly imaging low surface brightness structures presents a unique challenge as the features of interest are potentially millions of times fainter than the brightest objects in a typical image. 
This huge dynamic range coupled with systematic errors that are typically negligible or correctable (e.g., scattering in telescope optics, detector amplification) greatly impacts our ability to detect low surface brightness features.
The archetype of low surface brightness phenomena is the largest and faintest structure in the universe: the `cosmic web'. 
Cosmological simulations predict that on large scales dark matter collapses to form the cosmic web, taking on a foamlike structure that permeates the universe with galaxies forming at the nodes of this web. Gas flows along the filaments of the cosmic web (the intergalactic medium) into giant reservoirs of gas surrounding galaxies (the circumgalactic medium), then transitions into galaxies to fuel star formation. The gas in the cosmic web is nearly invisible, emitting mainly through fluorescence as the gas slowly cools.
The prospect of directly detecting the ultra-low surface brightness line emission from cosmic web has spurred the development of technological upgrades in astronomical instrumentation. In particular, there have been several recent and ongoing projects to build instruments that aim to directly observe line emission from the circumgalactic and intergalactic medium, such as the Cosmic Web Imager (CWI) at Palomar Observatory \cite{matu10} and the Keck Cosmic Web Imager (KCWI) at the W. M. Keck Observatory\cite{morr12,morr18}. 

Imaging the circumgalactic and intergalactic media remains very difficult due to the faintness of emission from the diffuse gas, but there has been recent success at high redshifts where rest-frame ultraviolet (UV) emission lines
are targeted by
integral field spectrometers such as CWI and KCWI, as well as the Multi-Unit Spectroscopic Explorer (MUSE) on the Very Large Telescope (VLT)\cite{baco09}. In the past decade, integral field units with spectroscopic capability have discovered an abundance of extragalactic gaseous structures, including gaseous bridges between galaxies and quasars\cite{arri19,lecl22}, enormous hundreds-of-kpc-sized gas clouds\cite{bord22,cai18}, and extensive gaseous haloes detected in a variety of emission lines including Ly$\alpha$, \textsc{[Oii]}, and Mg~\textsc{ii} \cite{lau22,burc21,rupk19,cai19}. These techniques work  well for medium to high redshift observations, where bright UV lines such as the Ly$\alpha$ $\lambda1216$ and Mg~\textsc{ii} $\lambda2796,2803$ emission lines are redshifted into the visible wavelength regime and where the $\lesssim1$ arcmin$^2$ fields of view of these instruments can cover an appreciable area around the targeted galaxies. In the local universe, though, the bright UV emission lines are not accessible from the ground and while visible wavelength emission lines such as H$\alpha$ and \textsc{[Oiii]} are accessible, they are an order of magnitude fainter, requiring sensitivity to surface brightnesses down to $\sim10^{-20}$~erg~cm$^{-2}$~s$^{-1}$~arcsec$^{-2}$ to detect\cite{lokh19}.

As we have shown in Ref.~\citenum{lokh19}, and discuss in detail below, it should be possible for an upgraded Dragonfly Telephoto Array\cite{abra14} (Dragonfly) with ultranarrowband imaging capability to reach this low surface brightness limit and directly detect visible wavelength line emission from gas in the circumgalactic medium\cite{lokh19,lokh20}. 
Dragonfly is a telescope specially designed for low surface brightness imaging.
Its design is based on the innovative concept of building an extremely fast refracting telescope using lenses instead of mirrors to reduce scattering of light in the optics and multiplexing together an array of high-end commercial lenses to synthesize a larger effective aperture.
Over the past few years, Dragonfly has made a series of groundbreaking discoveries, helping to rekindle the long-neglected study of the low surface brightness universe.  Dragonfly’s success in detecting low surface brightness stellar structures prompted the question of whether Dragonfly, with its unprecedented sensitivity to the diffuse stellar continuum, could be modified to detect the extremely faint line emission from the circumgalactic and intergalactic media.

The pathfinder \dslm~was built to test this concept. Its basic design and components were modeled off of the Dragonfly Telephoto Array with the addition of instrumentation to incorporate ultranarrowband imaging capability. The pathfinder \dslm~uses the Dragonfly Filter-Tilter instrumentation described in Ref.~\citenum{lokh20} to implement ultranarrowband filters on the telescope at the front of the optics. This is important for two main reasons: 1) to prevent degradation of the filter transmission profile that inevitably occurs when interference filters are placed in converging beams and 2) to incorporate a rotational mechanism to tilt the filter, which smoothly shifts the bandpass in wavelength space. Rotating an ultranarrowband filter by $\sim20^\circ$ enables it to target the same cosmological volume as a typical narrowband filter of bandwidth $\sim$ 10 nm. The pathfinder \dslm~was used to carry out an imaging survey on the M81 group of galaxies utilizing ultranarrowband filters selected to target H$\alpha$ and \textsc{[Nii]} emission. These observations functioned as both a proof-of-concept and a scientific survey of the group. The H$\alpha$ and \textsc{[Nii]} emission from the group was imaged separately with the same filters by rotating the ultranarrowband filters to two different angles with respect to the optical axis. The resulting data revealed many new gaseous features in the group, including a giant cloud of gas in the outskirts of the M82 galaxy\cite{lokh22}.

Here we describe the pathfinder \dslm~and the observing methods used to carry out the imaging survey. This includes several upgrades and additions to the observing methods and software used by the Dragonfly Telephoto Array, such as calibration of the Dragonfly Filter-Tilter and additional hardware for flat-field imaging. 
Finally, we compare the sensitivity of the Dragonfly Spectral Line Mapper with other observatories and instruments that are designed to detect similar features, and extrapolate on the final sensitivities expected to be reached by the full \dslm, which is under construction. The \dslm~will have 120 lenses, 40$\times$ the collected area of the pathfinder, including a specialized suite of filters to enable sub-percent calibration of the science frames.

This paper is one of three in a series in these proceedings on imaging the low surface brightness universe with distributed aperture telescopes and the Dragonfly Telephoto Array. Table \ref{tab:3papers} summarizes the topics covered by each paper.

\begin{table}[ht]
\caption{Content summary of the three papers in this series, with the topics covered by this paper in bold.} 
\label{tab:3papers}
\begin{center}       
\begin{tabular}{|l|l|} 

\hline
\rule[-1ex]{0pt}{3.5ex}  Distributed aperture telescope general concepts & Abraham et al. \\
\hline
\rule[-1ex]{0pt}{3.5ex}  Low surface brightness imaging challenges	 & Abraham et al. \\
\hline
\rule[-1ex]{0pt}{3.5ex}  Lessons learned from Dragonfly & Abraham et al.  \\
\hline
\rule[-1ex]{0pt}{3.5ex}  \textbf{Narrowband imaging concepts and methods} & This article  \\
\hline
\rule[-1ex]{0pt}{3.5ex}  \textbf{Narrowband imaging survey speed} & This article \\
\hline
\rule[-1ex]{0pt}{3.5ex}  \textbf{Dragonfly Spectral Line Mapper pathfinder results \& lessons learned} & This article \\
\hline
\rule[-1ex]{0pt}{3.5ex}  Dragonfly Spectral Line Mapper design  & Chen et al. \\
\hline
\rule[-1ex]{0pt}{3.5ex}  Dragonfly Spectral Line Mapper laboratory tests	 & Chen et al.\\
\hline
\rule[-1ex]{0pt}{3.5ex}  Dragonfly Spectral Line Mapper roadmap & Chen et al. \\
\hline
\end{tabular}
\end{center}
\end{table}

\section{the Dragonfly Spectral Line Mapper pathfinder}

\begin{figure}[t]
\centering
\includegraphics[width=0.3\linewidth]{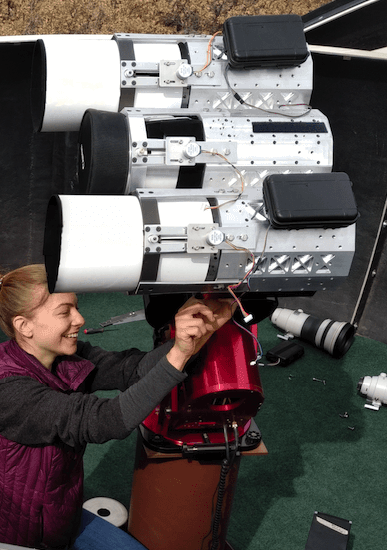}
\includegraphics[width=0.65\linewidth]{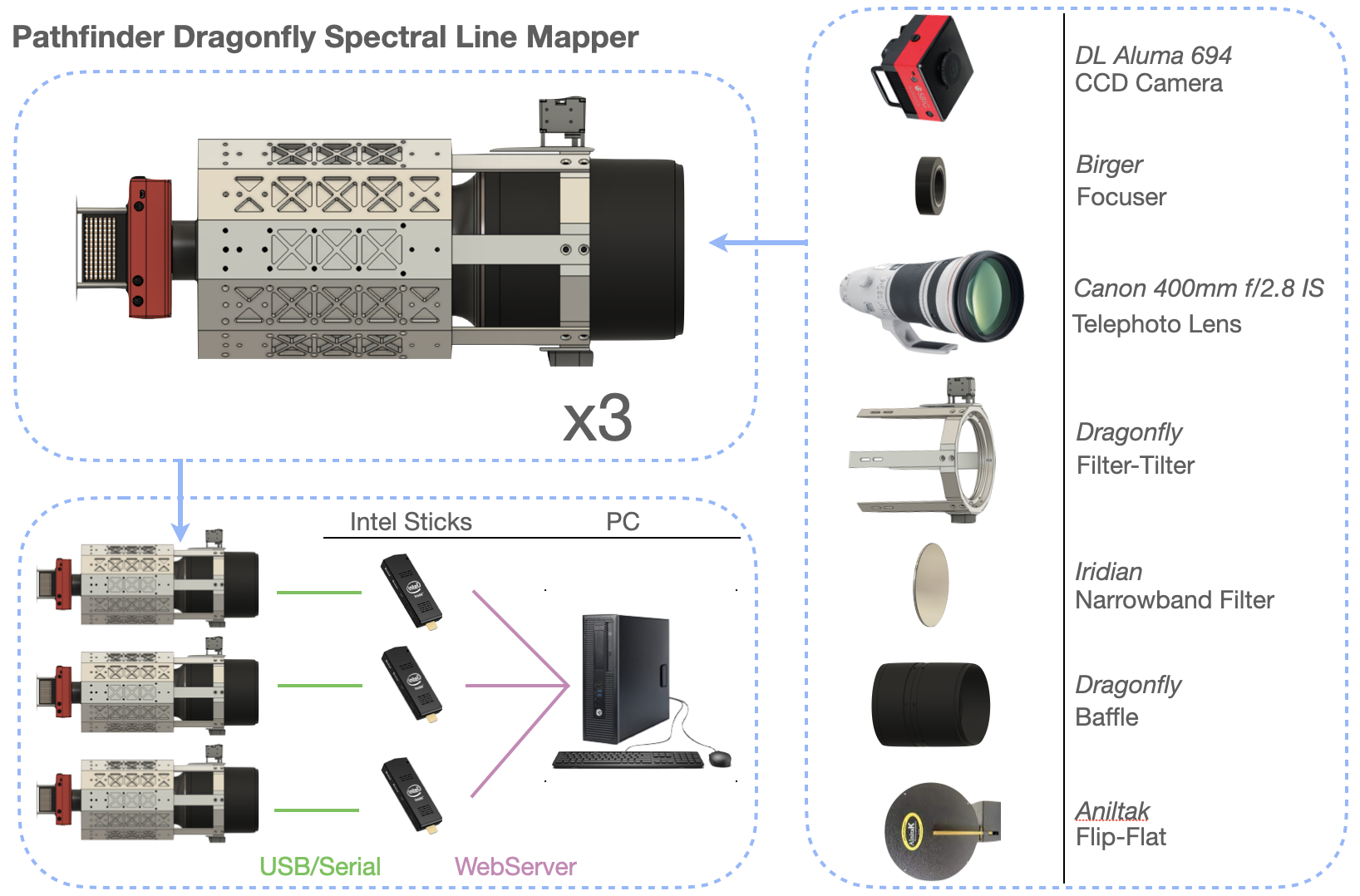}
\caption{ 
The pathfinder \dslm, on sky from the spring of 2019 to the fall of 2021 is pictured in the left panel. A schematic representation of the pathfinder is also shown, with the components and connections between computers and devices labeled.
}
\label{fig:deb1}
\end{figure}

The \dslmp~is a 3-lens version of the Dragonfly Telephoto Array with Dragonfly Filter-Tilter instrumentation that implements ultranarrow-bandpass imaging capability on the telescope (as described in Ref.~\citenum{lokh20}). The \dslmp~was built at New Mexico Skies Observatories in Mayhill, New Mexico, in the spring of 2019. Commissioning and science data were collected from 2019 through 2021. The pathfinder is composed of three separate and identical optical tube assemblies (OTAs; pictured in the left panel of Fig.~\ref{fig:deb1}). A schematic of the pathfinder is shown in the right panel of Fig.~\ref{fig:deb1}. Each OTA consists of a Canon telephoto 400mm $f$/2.8 IS lens with a 14.2 cm diameter aperture along with a Birger focuser, and a Diffraction Limited Aluma CCD694 camera with an angular scale of 2.45$''$ per pixel. This sensor has 2750 $\times$ 2200 pixels at 4.54 microns square, resulting in a 14.6 mm $\times$ 12.8 mm chip size and a 1$^\circ.4 \times 1^\circ.9$ field of view.
In addition, each OTA is equipped with Dragonfly Filter-Tilter instrumentation\cite{lokh20}, which holds a 152 mm diameter ultranarrowband filter at the entrance pupil of each OTA. The Dragonfly Filter-Tilter mechanically tilts the filter with respect to incident light, effectively changing the incidence angle and thus the central wavelength of the filter bandpass. Each Filter-Tilter is driven by a 28BJY-48 stepper motor with 2048 steps per revolution, controlled in a closed-loop feedback system by a Gravity 360 Degree Hall Angle Sensor. The filters are allowed to rotate from -20$^\circ$ to +20$^\circ$ around an axis perpendicular to the optical axis. Due to the properties of the narrowband interference filters, this shifts the central wavelength of the filter from its 0$^\circ$ angle of incidence value by up to $\approx$~8 nm. The filters used for these observations have a central wavelength of 659.9~nm in vacuum
and FWHM of 3~nm. See Ref.~\citenum{lokh20} for further details on the theory and instrument design of the Dragonfly Filter-Tilter. A custom designed baffle was built for each OTA to block stray light contamination from entering the optical path. The pathfinder instrument also includs electroluminescent flat field panels (commercially available Aniltak Flip-Flats) that were mounted at the front of each OTA to collect flat field images. The panels were used to collect flat field images during the night in between science exposures at the same pointing and filter rotation as used for the science frame collection.
The elements of the \dslmp~are summarized in Table~\ref{tab:dslmp} for quick reference.

\begin{table}[t]
\caption{Specifications of the pathfinder Dragonfly Spectral Line Mapper.} 
\label{tab:dslmp}
\begin{center}       
\begin{tabular}{|l|l|} 
\hline
\rule[-1ex]{0pt}{3.5ex}  \textbf{Parameter} & \textbf{Value} \\
\hline\hline
\rule[-1ex]{0pt}{3.5ex}  Effective aperture & 24.6 cm (14.3 cm $\times$ 3) \\
\hline
\rule[-1ex]{0pt}{3.5ex}  Effective focal length & 40 cm ($f$/1.6) \\
\hline
\rule[-1ex]{0pt}{3.5ex}  Field of view & 1$^\circ.4 \times 1^\circ.9$  \\
\hline
\rule[-1ex]{0pt}{3.5ex}  Camera & SBIG Aluma CCD694 \\
\hline
\rule[-1ex]{0pt}{3.5ex}  Detector & Sony ICX 694, 2750 $\times$ 2200 CCD  \\
\hline
\rule[-1ex]{0pt}{3.5ex}  Pixel size & 4.54 $\mu$m (2.45$''$) \\
\hline
\rule[-1ex]{0pt}{3.5ex}  Optics & Canon Telephoto 400mm f/2.8 IS II, \\ & Canon Telephoto 400mm f/2.8 IS II, \\ & Canon Telephoto 400mm f/2.8 IS III  \\
\hline
\rule[-1ex]{0pt}{3.5ex}  Filters & See Table~\ref{tab:filter}   \\
\hline
\rule[-1ex]{0pt}{3.5ex}  Mount & Paramount ME II   \\
\hline
\end{tabular}
\end{center}
\end{table} 

\begin{table}[t]
\caption{Specifications of the Iridian Spectral Technologies Inc. Dragonfly Filter.} 
\label{tab:filter}
\begin{center}       
\begin{tabular}{|l|l|} 
\hline
\rule[-1ex]{0pt}{3.5ex}  \textbf{Parameter} & \textbf{Value} \\
\hline\hline
\rule[-1ex]{0pt}{3.5ex}  Central Wavelength at 0$^\circ$ angle of incidence & 659.9 nm  \\
\hline
\rule[-1ex]{0pt}{3.5ex}  Bandwidth & 3 nm   \\
\hline
\rule[-1ex]{0pt}{3.5ex}  Out-of-band blocking & $<$OD4 200 - 1100 nm   \\
\hline
\rule[-1ex]{0pt}{3.5ex}  Clear Aperture Diameter & 152 mm   \\
\hline
\rule[-1ex]{0pt}{3.5ex}  Operating temperature & -10$^\circ$C to +20$^\circ$C   \\
\hline
\rule[-1ex]{0pt}{3.5ex}  Transmittance & Peak Tx $>$95\% with Tx $>$90\% for bandwidth $>$2.0 nm  \\
\hline
\end{tabular}
\end{center}
\end{table}

\subsection{Observing Software}

The pathfinder \dslm~used the observing software package written for the Dragonfly Telephoto Array\cite{abra14} with several additions to integrate the required ultranarrowband instrumentation. The Dragonfly observing software is designed to allow completely autonomous observing, through all manner of weather conditions. The observing software is adaptive; upon starting observations for the night, it checks before each science exposure whether it should continue or not (e.g., it will stop taking science images if the weather has reported cloudy or dome is closed). Depending on dome and sky conditions, it carries out the optimal task, such as a science exposure, dark frame exposure (due to cloudy weather conditions), or flat-field exposure (due to the rising of the sun). 
Dragonfly uses an Internet-of-Things framework to control each OTA of the telescope. In this framework, API commands are sent from a control PC to the OTAs, each of which has its own Intel Compute Stick. The observing commands are sent over web servers that run on each Intel Stick. Each Intel Stick is connected through USB or serial ports to the components of the OTA, allowing it to control the camera, focuser, Filter-Tilter and Flip-Flat for that specific OTA. The connections between the computers and devices are shown diagrammatically in the bottom-middle panel of Fig.~\ref{fig:deb1}.

The Dragonfly observing software worked well for controlling the pathfinder \dslm, but it was not a direct drop-in replacement. In particular, the Dragonfly Filter-Tilter control and flat-frame collection needed to be developed and integrated with the observing software. In addition, with the change from broadband to narrowband filters, several smaller changes, such as the adjustment of exposure times for focusing frames, were necessary. We detail these modifications and additions for the interested reader here. 

\begin{enumerate}
    \item \emph{Filter-Tilter Control Software.} In order to carry out observations on the pathfinder \dslm, the observing software needed to be updated to also include an API to control the Filter-Tilter rotation. The software to control the Filter-Tilters was integrated into the observing software to allow the filters to be automatically tilted to the correct rotation before each integration. This required adding RESTful API calls to the Dragonfly observing software. 
    Tilting the filters now requires just a single extra parameter when starting the observations and is fully automated during the night without needing any intervention from the observer. Additional software for carrying out automated Filter-Tilter calibration was also developed and is described in Section~\ref{sec:FTCalibration}.
 
    \item \emph{Flat field frame collection.} The observing software was updated to include gathering flat field frames using illumination panels (``Flip-Flats'') which are commercially available. Each lens on the pathfinder has its own Flip-Flat. During the night, flat frames at each observing tilt of the filter can be taken to allow an accurate calibration in the case of any flexure in the equipment. Observing software was added to carry out the flat frame collection automatically after each set of science exposures.

    \item \emph{Lens focusing.} The script for focusing the lenses was modified to allow the exposure time for the focusing images to be adjusted. Longer exposure times were necessary because of the significantly narrower bandwidth of the pathfinder \dslm~filters compared to the Dragonfly broadband filters.
\end{enumerate}

\subsection{Filter-Tilter Calibration Method}\label{sec:FTCalibration}

 \begin{figure*} [t]
  \begin{center}
  \begin{tabular}{c} 
  \includegraphics[width=0.99\linewidth]{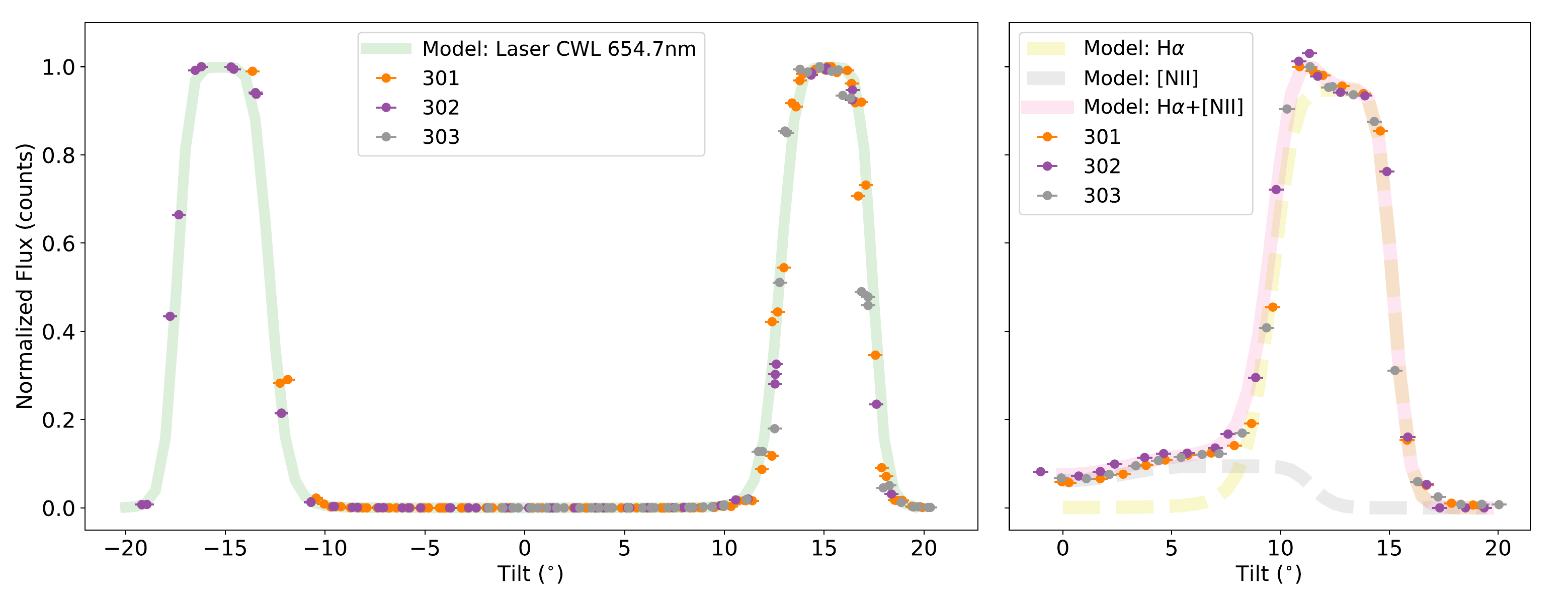}

  \end{tabular}
  \end{center}
  \caption[Laser flux and brightness of the planetary nebula NGC 6543 through the narrowband filters at a series of filter tilts, demonstrating the Filter-Tilter zeropoint tilt calibration method.]
  { \label{fig:calibration} 
    \emph{Left panel:} The laser flux through the narrowband filters measured for tilts between -20 and +20 degrees; the orange, purple and gray data points are measured from the units of the narrowband Dragonfly pathfinder independently.  A model of the laser throughput through the filters is shown as the green line.  The model and data are well-matched, providing a straight-forward calibration method by fitting the model to our data to find the zeropoint of the filter tilt.
  \emph{Right panel:} The measured brightness of the planetary nebula NGC 6543 at a series of filter tilts from 0 to 20 deg (shown for each unit of the Dragonfly narrowband set up as the orange, purple, and gray datapoints).  The measurements are compared to models of the planetary nebula throughput, taking into account H$\alpha$ and \textsc{[Nii]} emission from the planetary nebula.
}
\end{figure*} 

The operation of the pathfinder \dslm~relies on the ability to set the desired central wavelength of the filter by tilting the filters. In order to tilt to a specific wavelength on command, a calibration of the zeropoint of the filter tilt is required. In this Section, we describe the current calibration method.

The shift in the central wavelength of a filter is a smoothly varying monotonic function of the tilt of the filter and the angle of incidence of the light across the field-of-view can be easily modelled. This means that the calibration of the tilt of the filter is a straightforward process using a target radiating line emission (or at least quasi-monochromatic emission) at a wavelength within the range swept by the filter bandpass during tilting (i.e., within the 653.9 nm to 659.9 nm range scanned by the pathfinder \dslm~filters during tilting, which have a central wavelength of 659.9 nm). 
We used two different light sources for this calibration: 1) a laser with a central wavelength of 657.4 nm and bandwidth 0.61 nm (measured at room temperature) and 2) line emission from a planetary nebula.

The first method of calibration consists of taking images of the laser shining on a target at a distance of about 20 m from the telescope. The chosen laser is stable over periods of about 30 minutes, which is sufficient for laser calibration, which takes 5 to 10 minutes. However, the wavelength of the inexpensive laser used has a strong dependence on the temperature and, as the laser is mounted inside the telescope dome in the open air, there is a temperature difference of $\Delta \mathrm{T} \approx$~20~$^{\circ}$C between winter and summer nights. This results in a variation in the lasing wavelength of up to 5 nm for our observations. 
To mitigate this uncertainty in our calibration, we either use corrected laser specifications for the temperature during calibration or make the setup independent of the lasing wavelength by tilting to an angle of 20$^{\circ}$ in both directions and using the midpoint distance between the two peaks as the filter bandpass shifts through the lasing wavelength to calibrate the tilt.
A series of images of the laser-illuminated target were taken at different tilts, after which the images were analysed to determine the brightness of the reflected laser through the filter in each image.  These results were compared to a model of the throughput of the laser light as a function of filter tilt, which is shown in the left panel of Fig.~\ref{fig:calibration}. The laser flux was measured from a tilt of $-20^{\circ}$ to $+20^{\circ}$ to create the two peaks described above and by fitting the model to the data, the zeropoint of the filter tilt was determined.

The second method of calibration is identical to the first method, except that instead of using a target illuminated by a laser, an astronomical source of line emission is used. In the second panel of Fig.~\ref{fig:calibration} the results for imaging the planetary nebula NGC 6543 (the ``Cat's Eye'' Nebula) at a series of filter tilts from $0 - 20^{\circ}$ are shown.  In this case, the known wavelength of H$\alpha$ emission from the planetary nebula is used to model the transmission through the filter bandpass and fit to the calibration observations to determine the zeropoint of the filter.  
For this set of calibration observations, there is an excess of emission redward of the H$\alpha$ emission, which is attributed to \textsc{[Nii]} emission from NGC 6543. A combined model of H$\alpha$ and \textsc{[Nii]} emission at the redshift of NGC 6543 was created to determine a better fit to the data.  A relative fraction of 0.12 between the H$\alpha$ and the \textsc{[Nii]} emission yielded the best fit to the data. This method can be used to properly fit the observed emission curve for a variety of astronomical sources during calibration analysis.
 
\section{Ultra-low Surface Brightness Observations}\label{sec:noisy}

\begin{figure*} [t]
  \begin{center}
  \begin{tabular}{c} 
  \includegraphics[width=0.45\linewidth]{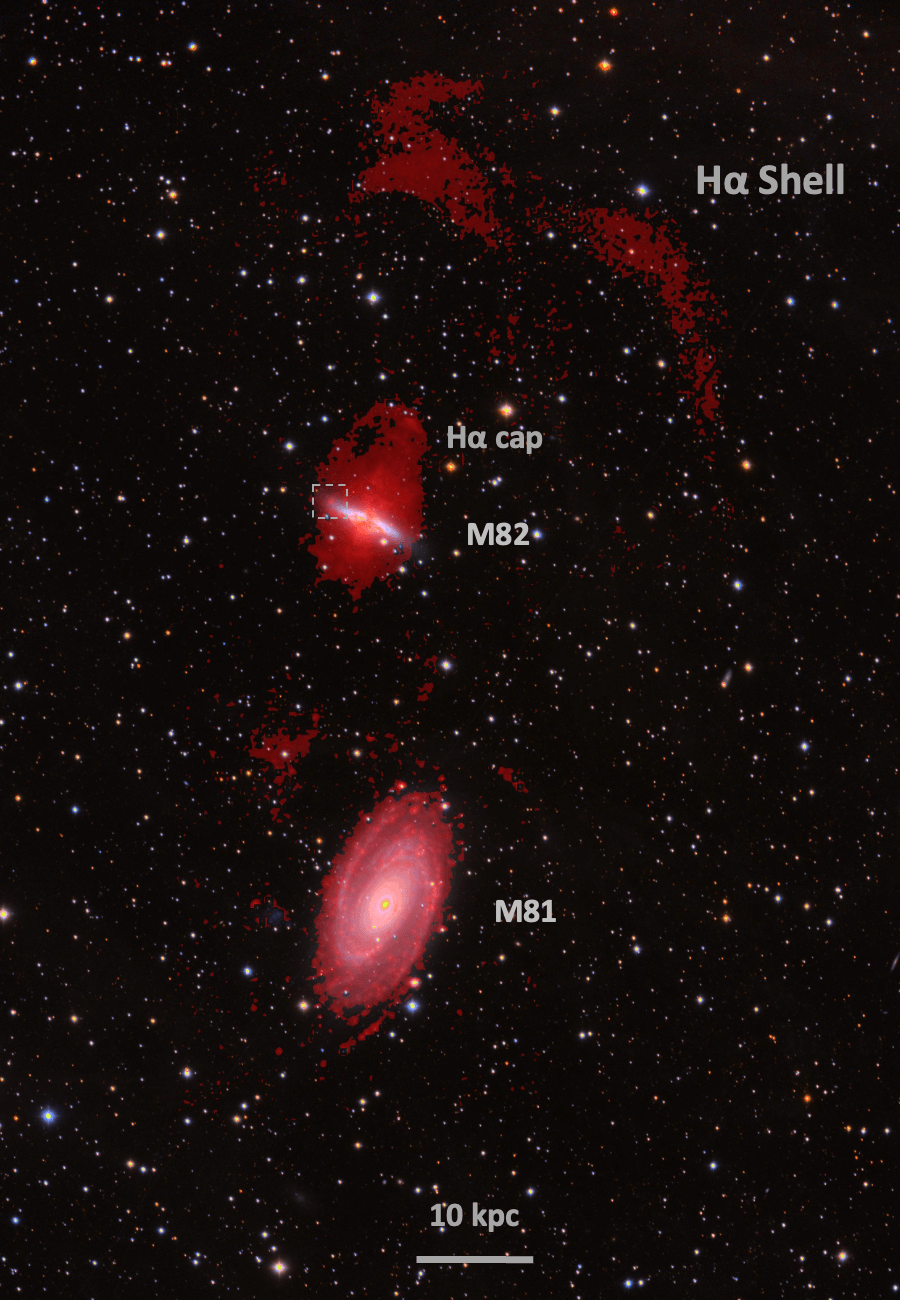}
  \end{tabular}
  \end{center}
  \caption{ \label{fig:hashell} 
    False colour image of M81 and M82 obtained with the Dragonfly Telephoto Array in $g$ and $r$ band imaging. Overlaid in red is H$\alpha$ data obtained with the 3-unit Dragonfly Spectral Line Mapper pathfinder. The H$\alpha$ imaging revealed a multitude of star formation complexes in M81 and the expansive M82 outflow that connects M82 to the ``H$\alpha$ cap'' above the disk\cite{devi99,lehn99}. In this imaging not only is the H$\alpha$ cap clearly visible but also a variety of new features including a tidal dwarf galaxy candidate at the edge of the M82 disk (Ref.~\citenum{pash21}; within the dashed gray box), H$\alpha$ emission from the tidal streamer appearing to connect the edge of the disk to the H$\alpha$ cap, and the most striking discovery of a giant cloud of gas seeming to hover above the M82 galaxy, the H$\alpha$ shell\cite{lokh22}.
}
\end{figure*}

An imaging campaign on the M81 group of galaxies was undertaken in the spring of 2020 with the pathfinder version of the \dslm. Narrowband imaging of H$\alpha$ and \textsc{[Nii]} emission from the galaxy group was collected from March 2020 to October 2021. These observations served as proof-of-concept imaging to determine the surface brightness limits of the instrument, as well as scientific imaging, the results of which are presented in Ref.~\citenum{lokh22}. 
The observations followed the Dragonfly automated observing model, where the telescope is set up every night for observing at the beginning of the night and the telescope carries out observations autonomously, adapting to changing weather conditions and pausing observations when necessary. In total, this resulted in 73 nights of data collection over the months of February to June 2020 and a total of 652 on-target science frames collected with individual exposure times of 1800 seconds. After the removal of frames with low image quality (e.g., due to poor focusing, poor guiding, poor seeing, etc.), the total exposure time with the 3-lens pathfinder that was included in the final science frames was 31.7 hours and 15.3 hours for the H$\alpha$ and \textsc{[Nii]} science images, respectively (see Ref.~\citenum{lokh20} for further details).
The data were taken with the filters at two different tilts: 12.5$^{\circ}$ to target the H$\alpha$~$\lambda6563$ emission line and 7$^{\circ}$ to target the \textsc{[Nii]}~$\lambda6583$ emission line. Tilting the filters smoothly shifts the filter central wavelength, and these two tilts shifted the filter central wavelengths to 656.3 nm and 683.5 nm, respectively. In addition, data were collected with the Dragonfly Telephoto Array (in the configuration most recently described in Ref.~\citenum{dani20}) yielding $r$ and $g$ band continuum images with exposure times of 12.5 minutes and 10 minutes, respectively. The final science images had a total field of view of $\sim$2$^\circ \times 3^\circ$ and reached an H$\alpha$ surface brightness limit of $\sim5 \times 10^{-19}$ erg cm$^{-2}$ s$^{-1}$ arcsec$^{-2}$ to the 3$\sigma$ level on a 4$'$ spatial scale. A composite $r$ and $g$ band image along with the high signal-to-noise regions of the pathfinder \dslm~H$\alpha$ data is shown in Fig.~\ref{fig:hashell}.

The deep, wide-field data confirmed previously discovered low surface brightness features in the field such as the M82 H$\alpha$ cap\cite{devi99,lehn99}, as well as several new features and extensions to already known features in the group. These included a secondary ridge of H$\alpha$ emission between the H$\alpha$ cap and the M82 galaxy, as well as emission filling the region between the M82 galaxy and the H$\alpha$ cap. In addition, H$\alpha$ emission from a tidal dwarf candidate and emission corresponding to the HI tidal streamer on the northeast side of the M82 disk were discovered (indicated by dashed box in Fig.~\ref{fig:hashell}; Ref.~\citenum{pash21}). The most striking result of this imaging campaign was the discovery of a colossal cloud of gas in the outskirts of the M82 galaxy with an extent of 0.8 deg (labelled in Fig.~\ref{fig:hashell}), which is analyzed further in Ref.~\citenum{lokh22}.

The main goal with this instrument was to reach the surface brightnesses required to directly image the ultra-faint line emission from gas in the surroundings of galaxies, and determine what improvements upon the instrument would be necessary before undertaking the construction of the full-scale \dslm.
In order to reach these low surface brightness limits, the calibration of our data needed to be undertaken with care. To ensure that our observations are sky noise limited, dark noise and flat-field corrections needed to be carried out to below 0.1\%. The data collection followed the Dragonfly model of collecting data regardless of sky conditions and then removing frames that are classified as ``bad''  during the data reduction pipeline. This automatically removed a significant fraction of the frames due to poor image quality. In addition to this data cut, we also inspected the frames by eye to search for and remove frames with contamination that were not caught by the automated pipeline. Together, this resulted in using only 279 out of the 652 science frames collected to create the final science images. In the following Sections, we delve into the analysis of the main sources of noise in the narrowband data and the methods for their removal. We identify several improvements in hardware and calibration frame collection that will be incorporated in the full-scale \dslm~design and operation.

\subsection{Dark Frame Stability}

 \begin{figure*} [t]
  \begin{center}
  \begin{tabular}{c} 
  \includegraphics[width=0.99\linewidth]{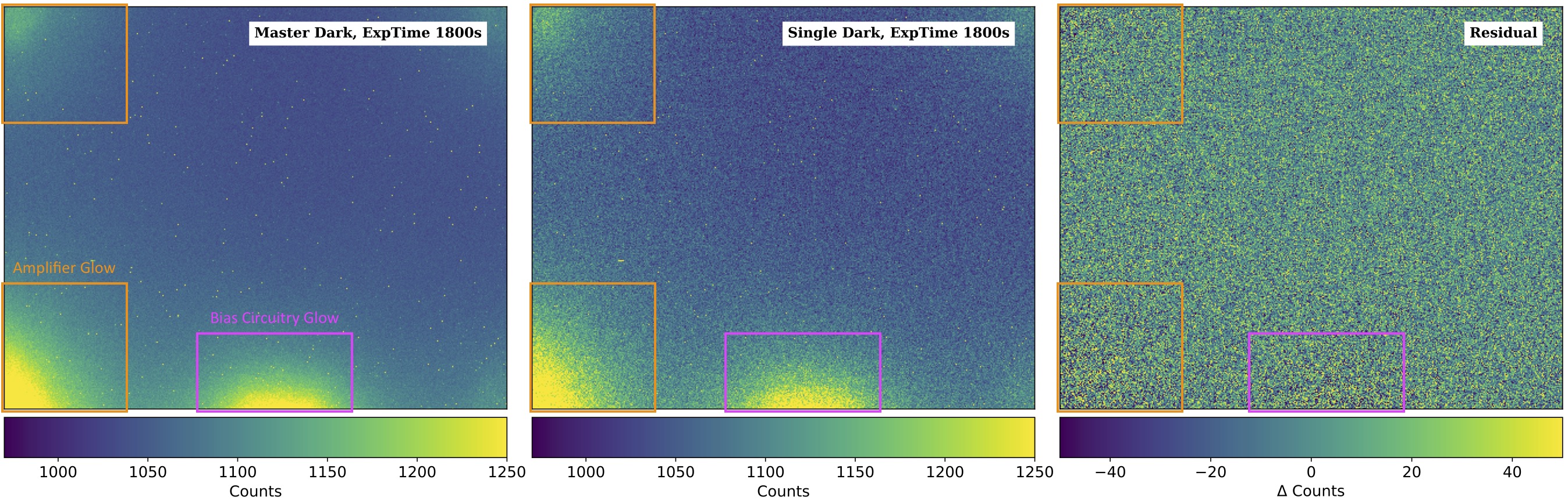}
    \end{tabular}
  \end{center}
  \caption{ \label{fig:darks} 
    A master dark formed from the combination of 10 1800s dark frames is shown in the left panel, a single dark frame of 1800s is shown in the middle panel, and the residual from the subtraction of the master dark from the single dark frame is shown in the right panel. There is significant amplifier glow and bias circuitry glow visible in the dark frames (locations most affected are indicated on the frames).
}
\end{figure*} 
 \begin{figure*} [t]
  \begin{center}
  \begin{tabular}{c} 
  \includegraphics[width=0.99\linewidth]{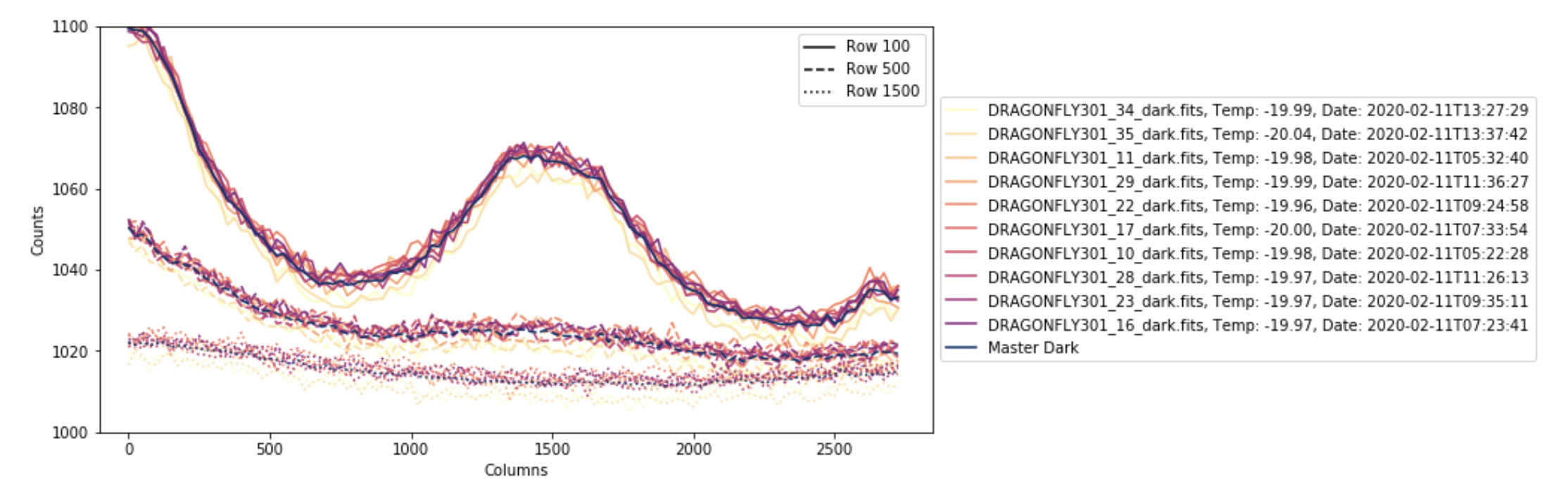}
    \end{tabular}
  \end{center}
  \caption{ \label{fig:darks2} 
    Comparison of the counts in rows 100, 500, and 1500 (averaged over the surrounding 10 rows, and averaged over 25 columns to reduce noise) in 10 1800s dark exposures taken on the same night along and the master dark made from the combination of these 10 dark frames. There is a shift in the bias level between darks, which is nonlinear in the regions affected by bias circuitry and amplifier glow (e.g., row 100).
}
\end{figure*} 

 \begin{figure*} [t]
  \begin{center}
  \begin{tabular}{c} 
  \includegraphics[width=0.8\linewidth]{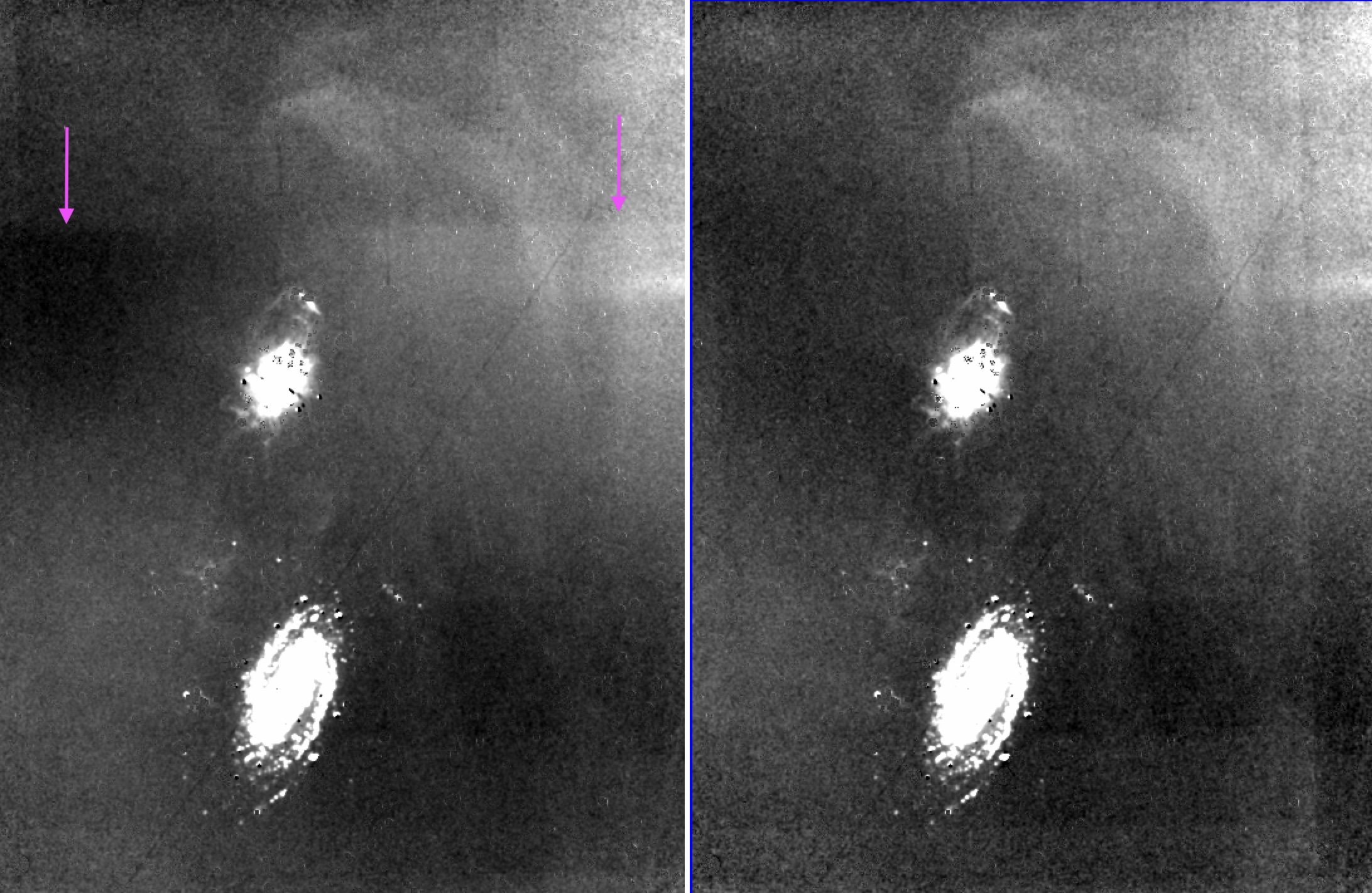}
    \end{tabular}
  \end{center}
  \caption{ \label{fig:darks3} 
    Comparison of H$\alpha$ science exposure stack with and without removing frames with poor dark frame subtraction. Including the frames with poor dark subtraction (left panel) results in large variation with a cutoff along a specific declination. Removing those frames and redoing the science exposure stack (right panel) removes that linear feature, but also results in lower signal-to-noise throughout the frame. This effect is partly due to repeated dither positions used during exposures on different nights which ended up aligning the regions with dark noise in the science image.
}
\end{figure*} 

The Aluma CCD694 cameras used with the \dslm~pathfinder were selected for their low dark current (0.025 electron s$^{-1}$ pixel$^{-1}$) and readout noise (4 electron RMS). 
In half an hour dark frame exposures, it was observed that there were additional significant sources of noise in the images above the dark current and readout noise: glow from amplifier and bias circuitry (e.g., Figs~\ref{fig:darks}~and~\ref{fig:darks2}). While the dark current and readout noise met the requirements for the low surface brightness limits in the data, the noise from the amplifier and bias circuitry proved difficult to remove due to their fluctuation over time. To quantify this variation, we collected a series of half an hour dark frames over each night for one week and a set of dark frames spread out over the time frame of a month. While the background noise level in the regions of the frame unaffected by glow from the amplifier and bias circuitry stayed consistent with the Poisson noise of the dark current (RMS$\sim$30 counts in the raw images), the amplifier and bias circuitry noise varied considerably. 

The amplifier glow and bias circuitry glow is visible in the dark images displayed in Fig.~\ref{fig:darks}, which shows a master dark formed out of 10 dark frames with exposure times of 1800s on the left panel, a single dark with an exposure time of 1800s in the middle panel, and the residual from the subtraction of the master dark from the single dark in the right panel. This residual is a replica of the dark noise that would remain after dark frame subtraction in a science exposure of 1800s, representing the combination of noise from the dark current and readout. An increase in noise is visible in the areas affected by the amplifier and bias circuitry. 
Fig.~\ref{fig:darks2} shows averages over rows in 10 dark frames taken during one night, as well as the master dark. It is apparent that the underlying bias level shifts through the night, so the master dark is not representative of the complete sample of dark frames. The shift in the bias level adds a noise floor to the data which is removed during the sky modeling and subtraction step of the Dragonfly pipeline. Any nonlinearity in the noise floor such as the amplifier glow and bias circuitry glow remains in the science frame, though, as they are not fit and subtracted during the sky modeling step.
This added a systematic noise source into the science frames, creating an artificial low surface brightness feature within the final science frames. This feature was apparent in the final science image despite the $\sim15'$ dithering of the telescope between science exposures and $\sim15'$ offset in pointing between the three OTAs. In order to remove this noise source, the data reduction pipeline was modified with an additional step of cropping the raw images to remove the bottom physical section of the image where the amplifier and bias circuitry glow showed up. This successfully removed the false feature caused by the systematic noise in the science frame. This is shown in Fig.~\ref{fig:darks3}, which displays H$\alpha$ science image stacks with and without removing frames with poor dark frame subtraction. Including the frames with poor dark subtraction (left panel) results in large variation with a cutoff along a specific declination. Removing those frames and redoing the science exposure stack (right panel) removes that linear feature, but also results in lower signal-to-noise throughout the frame. This effect is partly due to repeated dither positions used during exposures on different nights which ended up aligning the regions with dark noise perfectly in the science image. We will be using randomized dither patterns with the full \dslm~to avoid this effect, which is magnified as the exposure times increase in length. Newer Aluma CCD694 cameras use modified electronics and firmware, and the amplifier and bias circuitry glow have been significantly reduced since this data was collected (see our companion paper, Chen, S. et al. 2022, for further details).

\subsection{Flat Fielding Stability}

\begin{figure}[t]
	\centering
    \includegraphics[width=1\textwidth]{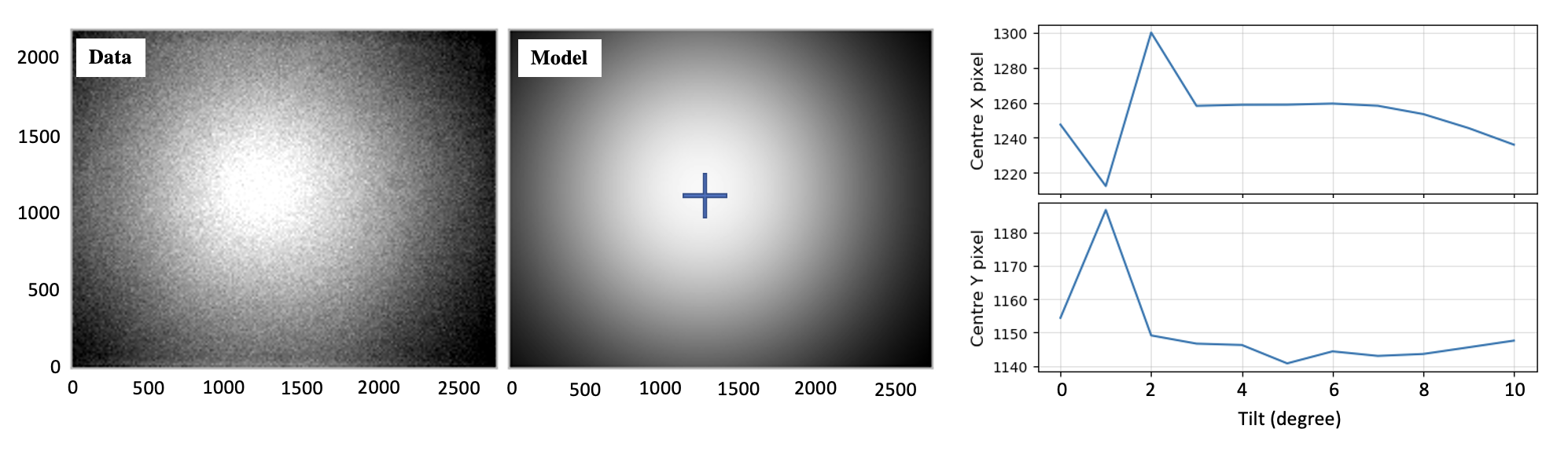}
    \caption{The left plot shows a calibrated and stacked flat taken by one of the pathfinder \dslm~OTAs at a filter tilt of 0 degrees using the electroluminescent flat-fielding panel. The middle plot shows the $\cos^4$ vignetting model (Equation \ref{cos4}) fitted to the data. The center x-pixel, $x_c$, and center y-pixel, $y_c$, for each flat as a function of filter angle are shown in the right panel, where the center position is calculated from the $\cos^4$ vignetting model.
    }  
    \label{fig:modelflats}
\end{figure}

Flat-fields for the \dslm~pathfinder were collected using electroluminescent flat-fielding panels (Alnitak Flip-Flats). This allowed us to collect flats at a series of different tilts. To investigate the effects of different filter tilt angles on the flat fielding, flats were taken at tilt angles from 0 to 10 degrees, in increments of 1 degree. Five flats were taken at each angle, and combined to make a master flat at each angle. We modeled the bright center and dimmer edges as $\cos^4$ vignetting for the master flat at each angle; an example is shown in Fig.~\ref{fig:modelflats} for a filter rotated to an angle of 0 deg. The $\cos^4$ vignetting is given by the following function:
\begin{equation}
    z = A \cos^4\left(\sqrt{\left(\frac{x - x_c}{a}\right)^2 + \left(\frac{y-y_c}{b}\right)^2}\right) + B
    \label{cos4}
\end{equation}
where z is the pixel value, $x_c$ is the center x-pixel, $y_c$ is the center y-pixel, $a$ and $b$ are the semi minor/major axes of the ellipse, and  $A$ is the scaling amplitude, and $B$ is the offset. For simplicity, $a=b$, as after fitting the data with both $a$ and $b$ variable we find $a \approx b$, differing by about 1-3 pixels.



The center point of the vignetting moved $\sim10-100$ pixels depending on the tilt of the filter. The right panel of Fig.~\ref{fig:modelflats} shows the variation of the  center pixel location in the fitted model as a function of the tilt of the filter. At larger angles the center pixel $(x_c, y_c)$ moves “left and up”. 
The vignetting depends on the tilt of the filter, making it necessary to collect flats at the same tilt angles as those used during Dragonfly science observations to carry out a robust calibration from the flat field images. The nonlinearity of the shift in the position of the center pixel with filter angle may be due to the shift in the shape of the spectral response across the field of view which is symmetric at a filter tilt of $0^\circ$ then becomes a linear gradient across the field of view at a filter of tilt of $\approx3^\circ$ (see Figure 6 of Ref.~\citenum{lokh20}). 

In addition, we were careful to collect flats while the telescope was at the same pointing as the science frames were collected. By utilizing the Flip-Flats, we were able to collect flats throughout the night, before and after each science exposure. This is necessary due to flexure in the optical tube assembly, in particular caused by shifting of the image stabilisation lens in the Canon telephoto lenses and flexure in the connection between the lens and focuser-camera assembly. Similar to that from varying tilts, this flexure resulted in a shift in the centroid of the flat-field image. While the centroid shift was very small (on order of a few pixels), the resulting additional source of noise was significant.


\subsection{Stray Light Contamination}

 \begin{figure*} [t]
  \begin{center}
  \begin{tabular}{c} 
  \includegraphics[width=0.8\linewidth]{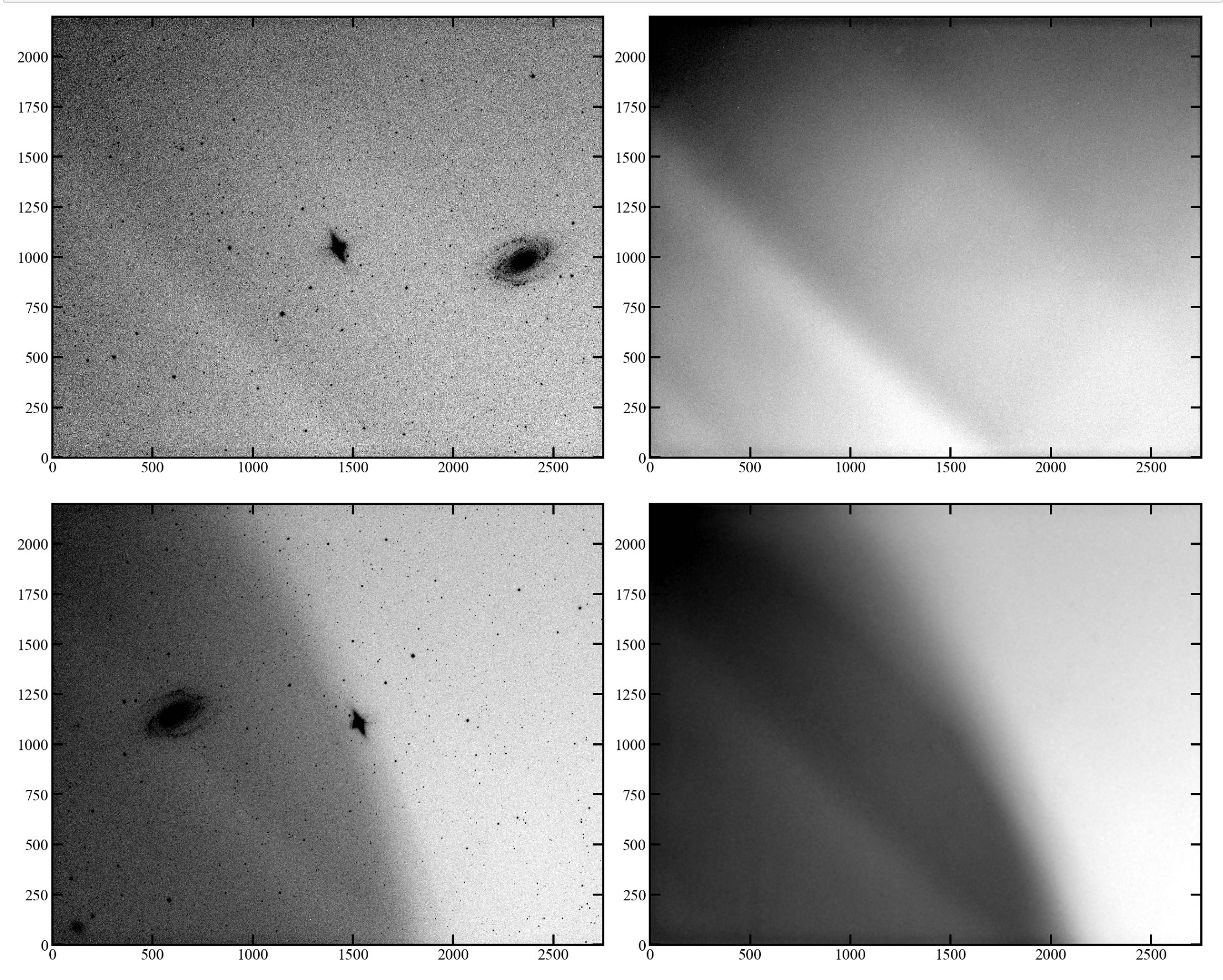}
    \end{tabular}
  \end{center}
  \caption{ \label{fig:lightstructure} 
    Many of the frames collected by the pathfinder \dslm~where contaminated by light structures. The left column of images shows two prominent examples that together affected $\sim100$ data frames. In the right column is models of the background sky which shows two distinct types of light features that appeared for some time in the data.
}
\end{figure*} 

In this narrowband setup, we are extremely sensitive to any sources of scattered light or light leakage through the optical tube assembly onto the detector. During the observations, we noticed that two strong types of contaminating light structures appeared in a subset of the data frames on different nights. Two example images showing data contaminated by these two light structures are shown in the left column of Fig.~\ref{fig:lightstructure}. Models of the background light structure are shown in the right column of Fig.~\ref{fig:lightstructure}. All three of the OTAs in the pathfinder \dslm~were affected by light contamination to a low level that usually showed up as a gradient in the images. The structures shown in Fig.~\ref{fig:lightstructure} are unique in that they are extremely prominent (at least an order of magnitude larger than typical light contamination in the images) and appeared solely in data taken with the OTA that had the latest generation of the telephoto lens, the 400mm f/2.8 IS III version. After further investigation, we determined that at least part of this light contamination was due to faulty baffling of the Filter-Tilter that allowed light to leak into the front of the optics. Indeed, the situation for one of these prominent light structures (shown in the bottom panels of Fig.~\ref{fig:lightstructure}) resolved itself without any intervention from the science team, so perhaps it was a hardware issue that was fixed as part of routine servicing of the telescope by the observatory staff. To address the other large source of light contamination (shown in the top panels of Fig~\ref{fig:lightstructure}), we carried out an extensive investigation to attempt to pinpoint the location of the light leak inside the optics. This included switching out every single component of our optical array. We finally concluded that it was the telephoto lens itself that was the culprit for causing the light contamination and we hypothesize that this light leakage is due to removal of shielding material in the telephoto lens allowing infrared emission to penetrate the lens and enter the optics, where it was then incident upon the detector. To test this theory, we inserted a UV-IR blocking filter into the available slot on the telephoto lens, which is at the back of the lens in front of the camera. This greatly reduced the light contamination in the optics and we were able to recover a clean signal in the image. 





\section{COSMIC WEB DETECTORS}

\begin{table}[t]
\caption{Instrument surface brightness limit and field of view comparisons for the Multi-Unit Spectroscopic Explorer (MUSE) on ESO's Very Large Telescope (VLT), the Cosmic Web Imager (CWI) at the Palomar Observatory, the Keck Cosmic Web Imager (KCWI) at the Keck Observatory, MegaCam imager at the Canada-France-Hawaii Telescope (CFHT), the Burrell Schmidt Telescope at Kitt Peak National Observatory, and the pathfinder and full-scale \dslm s (DSLM).} 
\label{tab:instruments}
\begin{center}       
\begin{tabular}{|l|l|l|l|l|l|} 
\hline
\rule[-1ex]{0pt}{3.5ex}  Instrument & Surface Brightness Limit & Spatial Scale & Exposure & FOV & Refs \\
\rule[-1ex]{0pt}{3.5ex}   & (erg cm$^{-2}$ s$^{-1}$ arcsec$^{-2}$) &  & Time (s) &  &  \\ 

\hline\hline
\rule[-1ex]{0pt}{3.5ex}  VLT/MUSE       &  $2.4\times10^{-19}$ at $3\sigma$ per $3.75\AA$ & $10''$       &  1 hr      & $60''\times60''$ & \citenum{sanc22} \\
\hline
\rule[-1ex]{0pt}{3.5ex}  Palomar/CWI    &  $1.3\times10^{-19}$ at $1\sigma$ per 4$\AA$ & $10''-15''$ &   11 hr    &  $60''\times40''$ & \citenum{mart14,mart14b} \\
\hline
\rule[-1ex]{0pt}{3.5ex}  Keck/KCWI      &   $4.8\times10^{-19}$ at $1\sigma$ per 5$\AA$  &  $1''$     &    3.5 hr   &  $33.1''\times20.4''$ & \citenum{burc21} \\
\hline
\rule[-1ex]{0pt}{3.5ex}  CFHT/MegaCam   &   $2\times10^{-18}$ at $1\sigma$   &  $3''$     & 2 hr      & $40'\times30'$ & \citenum{bose21} \\
\hline
\rule[-1ex]{0pt}{3.5ex}  Kitt Peak/   &   $3.09\times10^{-18}$ at $1\sigma$   &  $2'$     &   900s    & $1^{\circ}\times1^{\circ}$ & \citenum{dona95} \\
\rule[-1ex]{0pt}{3.5ex}  Burrell Schmidt & & & & & \\
\hline
\rule[-1ex]{0pt}{3.5ex}  Pathfinder DSLM &   $5\times10^{-19}$ at $3\sigma$   &  $4'$     &  32 hr   & $1^{\circ}.4\times1^{\circ}.9$ & \citenum{lokh22} \\
\hline
\rule[-1ex]{0pt}{3.5ex}  DSLM &   $1\times10^{-19}$ at $3\sigma$   &  $4'$     &  7 hr   & $1^{\circ}.4\times1^{\circ}.9$ & \citenum{lokh22}, this work \\
\hline
\end{tabular}
\end{center}
\end{table}

In this Section, we return to the earlier discussion of the telescopes and instruments being used to investigate the intergalactic and circumgalactic media of galaxies. There are now a large number of experiments targeting these media in the universe, ranging from local observations of galaxies to high redshift observations of quasars. In this Section, we provide a comparison of these observational techniques and the surface brightness limits they have reached. Each technique has relative strengths and weaknesses, so a side-by-side comparison allows one to choose the optimal instrument for specific observations. In Table~\ref{tab:instruments} we list the fields of view of several instruments that have been used to investigate the circumgalactic and/or intergalactic media of galaxies at different ranges of redshift, along with the surface brightness limits reached by these instruments when undertaking these studies. 
In addition to instruments with measured surface brightness limits, we have included an estimate of the surface brightness limits that will be observed with the full \dslm. This estimate was calculated by scaling the pathfinder \dslm~surface brightness limit by both increasing the aperture of the telescope (40$\times$ the size of the pathfinder) and reducing the pathfinder's bandpass (0.8 nm compared to the 3 nm bandpass of the pathfinder instrument). We predict that the \dslm~will reach a limiting surface brightness of $1\times10^{-19}$ erg cm$^{-2}$ s$^{-1}$ arcsec$^{-2}$ for a $3\sigma$ detection over a spatial scale of 4$'$ in 7 hours.

\begin{figure*} [t]
  \begin{center}
  \begin{tabular}{c} 
  \includegraphics[width=0.8\linewidth]{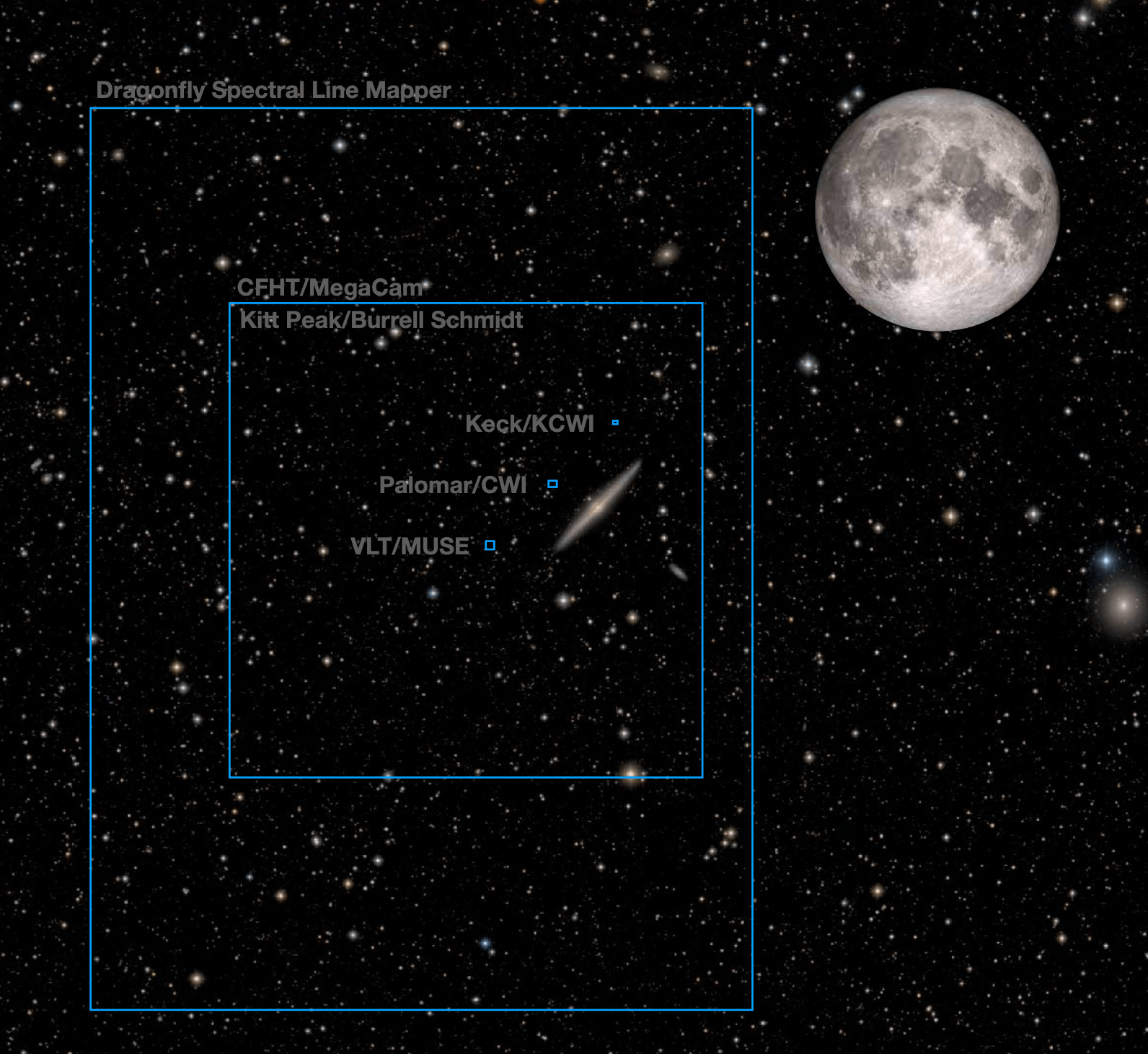}
    \end{tabular}
  \end{center}
  \caption{ \label{fig:fovs} 
    The fields of view of several instruments used to map the cosmic web as projected onto a false colour image of NGC 4565 (the Needle Galaxy) taken by the Dragonfly Telephoto Array\cite{gilh20}. The moon is shown for scale.
}
\end{figure*}

The fields of view of the instruments listed in Table~\ref{tab:instruments} are displayed in Fig.~\ref{fig:fovs}, projected on an image of the NGC 4565 galaxy (NGC 4565 has a radial distance of D $\approx13$ Mpc; Ref.~\citenum{tull16}). This image demonstrates the complementary nature of the various instruments when targeting structures in the nearby Universe (D $<$ 50 Mpc). Integral field units (IFUs) on 8-m class telescopes carry out high angular resolution studies, which are suited to smaller scale structures, while the wide-field imagers such as Dragonfly and CFHT MegaCam are suited to studying larger scale structures such as the galaxy environment. 

 \begin{figure*} [t]
  \begin{center}
  \begin{tabular}{c} 
  \includegraphics[width=0.7\linewidth]{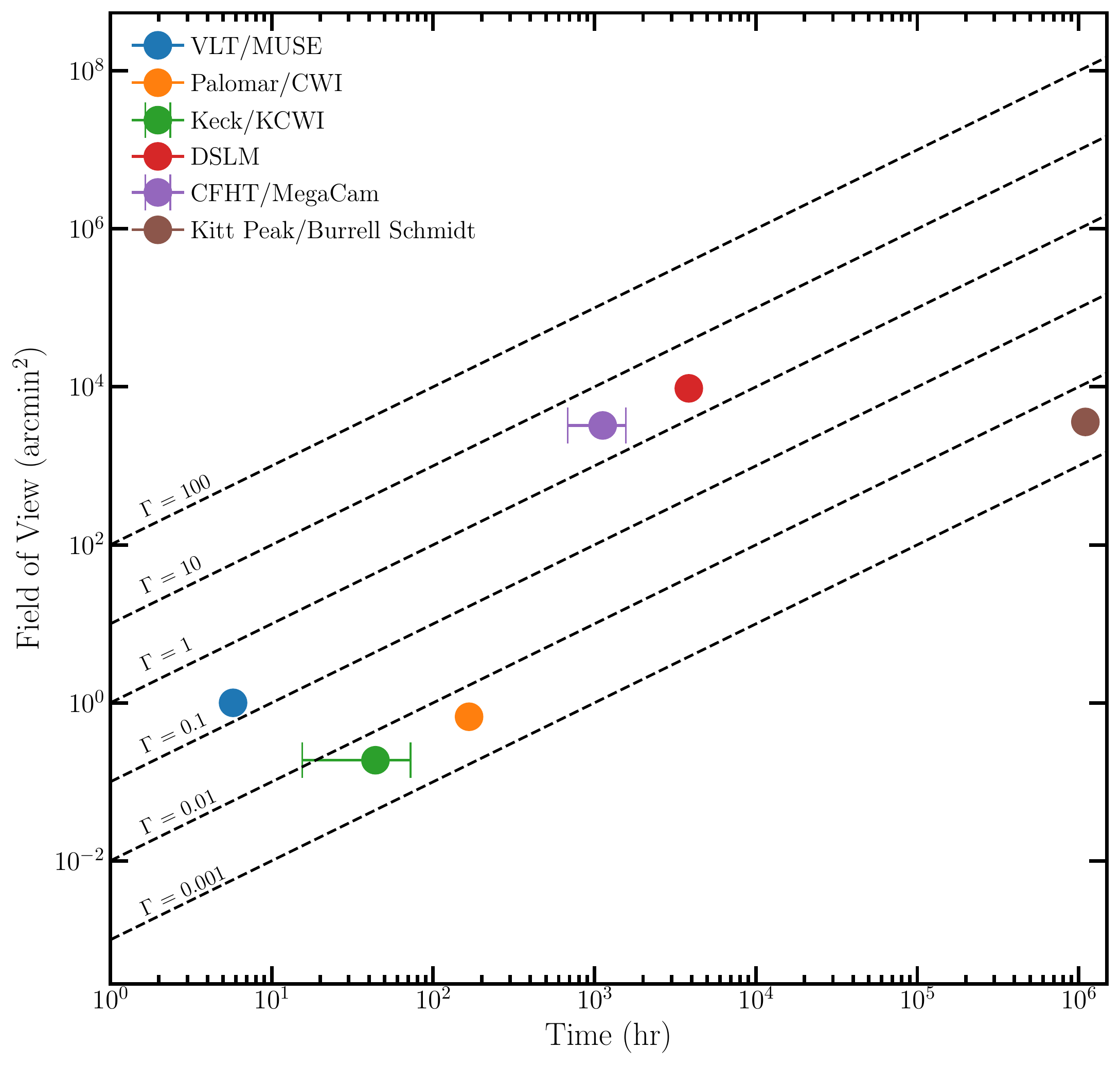}
    \end{tabular}
  \end{center}
  \caption{ \label{fig:surveyspeed} 
    A comparison of the fields of view of select instruments against the exposure time required to reach a sensitivity of 10$^{-19}$ erg cm$^{-2}$ s$^{-1}$ arcsec$^{-2}$ with a 3$\sigma$ detection. The surface brightness limits are calculated over a spatial scale of 10 arcsec$^2$ for the exposure time estimates. The surface brightness estimates were either taken directly from the literature or scaled from values in Table~\ref{tab:instruments}. We scaled these values empirically using relative numbers from instruments where the surface brightness limits have been calculated at different scales (e.g., Refs.~\citenum{baco21} and \citenum{sanc22} provide two different estimates of the sky background observed by the MUSE instrument). This introduces error into this estimate, particularly for the CFHT/MegaCam and Keck/KCWI estimates.
    Lines of constant survey speed are shown for comparison, where a higher survey speed value corresponds to a more efficient survey instrument.
}
\end{figure*}

One tool to measure the survey efficiency of an instrument is to calculate the sky survey rate, $\Gamma$, which is defined as the ratio of the observed sky area to the time needed to reach the desired sensitivity\cite{tere11}. We evaluate the instruments using this scale in Fig.~\ref{fig:surveyspeed}, where we have plotted the fields of view of the instruments against the time required to reach a sensitivity of $1\times10^{-19}$ erg cm$^{-2}$ s$^{-1}$ arcsec$^{-2}$ with a 3$\sigma$ detection. The surface brightness limits for the values in Fig.~\ref{fig:surveyspeed} are calculated over a spatial scale of 10 arcsec$^2$ and scaled from the values in Table~\ref{tab:instruments} where necessary. Scaling the detection limits from N$\sigma$ to 3$\sigma$ is straightforward as the surface brightness limit is calculated from the RMS fluctuations in the sky background, so a simple scaling of 3/N can be implemented. The required exposure times were scaled from the values in Table~\ref{tab:instruments} by assuming that the signal-to-noise ratio scales with the square root of the integration time. Creating an equivalent comparison between the spatial scale is less straightforward, and in this case, the values were scaled empirically using relative amplitudes from instruments where surface brightness limits calculated at different scales exists (e.g., Refs.~\citenum{baco21} and \citenum{sanc22} provide two different estimates of the sky background observed by the MUSE instrument). This introduces error into this estimate, which is encompassed by the error bars in the plot.

Lines of constant survey rate are plotted in Fig.~\ref{fig:surveyspeed} to aid the comparison of the survey efficiency of the instruments. As one may expect, the wide-field instruments have the highest survey efficiency due to their large fields of view. It is necessary that this large field of view is accompanied with sensitivity, though, which separates CFHT/MegaCam and the \dslm~from the Kitt Peak/Burrell Schmidt instrument. This plot also shows that when large fields of view are not important, the IFUs on the larger telescopes are more efficient for observations, due to the lower exposure times required.
There are limits to this comparison, which include that it does not take into account the spectral range of the IFUs, which adds the third spatial dimension into the survey, changing it from a survey area into a survey volume. Thus, this specific comparison is useful for considering single targets or objects within some fixed cosmological volume. Including the spectral range of the instruments would boost the survey efficiency of the IFUS. Another limitation of this comparison is the simplicity in calculating the limiting surface brightnesses. Not only is it approximate due to empirical scaling of sensitivities, but it also does not take into account systematic errors in surface brightness measurements. As was discussed earlier, Dragonfly's strength is the removal of systematic sources of noise in low surface brightness imaging, which isn't considered in this simple surface brightness limit estimate, which is based on the random noise in the images.



\section{Summary}

We have developed an upgrade to the Dragonfly Telephoto Array to implement ultranarrow bandpass imaging capability on the telescope with the goal of targeting the line emission from the circumgalactic medium around nearby galaxies, and the brightest pockets of the cosmic web in the local Universe. The Dragonfly Filter-Tilter\cite{lokh20} was developed to incorporate ultranarrowband filters on Dragonfly, mounting the filters at the front of the optics to avoid the degradation of the filter transmission profile and to enable the rotation of the filter with respect to the optical axis. We built the \dslm~pathfinder, which is a 3-unit version of Dragonfly with the Filter-Tilter instrumentation, to test this concept. We carried out an imaging on the M81 group of galaxies with the pathfinder reaching surface brightness limits comparable to those reached by state-of-the-art instruments on large optical telescopes. The low surface brightness levels reached by the pathfinder \dslm~forecast what the full \dslm~will reach. With narrower filter bandpasses and 40$\times$ more collecting area, we predict that the \dslm~will reach surface brightness limits of $1\times10^{-19}$ erg cm$^{-2}$ s$^{-1}$ arcsec$^{-2}$ on scales of 4$'$ in 7 hours. The \dslm~is under construction in a phased timeline, deploying 10 to 30 units at a time in stages. There are currently 10 units of the \dslm~on sky, which are being used to carry out commissioning and filter testing (see Chen, S., et al. 2022 for further details). This number will increase to 60 units in the fall of 2022, and to the final number of 120 units in the summer of 2023.

\appendix    

\acknowledgments 
 
We are very grateful to the staff at New Mexico Skies Observatories, without whom this work couldn’t have been carried out. This work is supported by a Canadian Foundation for Innovation (CFI) grant. We are thankful for contributions from the Dunlap Institute (funded through an endowment established by the David Dunlap family and the University of Toronto), the Natural Sciences and Engineering Research Council of Canada (NSERC), and the National Science Foundation (NSF), without which this research would not have been possible.

\bibliography{references} 
\bibliographystyle{spiebib} 

\end{document}